\documentclass[12pt]{article}

\usepackage{graphicx}
\usepackage{slashed}
\begin{document}

\title{Rotating Away Proton Decay in Flipped Unification}
\author{{\bf S.M. Barr} \\
Department of Physics and Astronomy and \\
The Bartol Research Institute \\ University of Delaware \\
Newark, Delaware 19716} \maketitle

\begin{abstract}
It is shown that by a simple extension of the fermion sector of flipped $SU(5)$ models and other flipped models proton decay coming from dimension-6 operators can be suppressed by fermion mixing angles by an arbitrary amount in a natural way. 
\end{abstract}

In unified theories there are mixing angles that describe how the Standard Model (SM)
quark and lepton multiplets are arranged within the multiplets of the unified
group. These unification mixing angles come into proton decay amplitudes \cite{DeRGG}, and if some of these angles are small, it is possible that proton decay could be greatly suppressed.

It is an interesting question whether proton decay might be {\it completely} ``rotated away"
in unified theories by a suitable choice of these mixing angles. For such a choice of angles, what would happen is that the lighter quarks and leptons
would be partnered within the irreducible multiplets of the unification group with fermions that are too heavy to appear in the final state of proton decay (which we will call ``heavy fermions").
The tree-level amplitudes for proton decay would therefore vanish. (At higher-loop
level, these heavier fermions might convert into lighter flavors, but the loop
factors of $(16 \pi^2)^{-2}$, not to mention suppressions of the GIM type, would then effectively kill proton decay anyway.)

There are two ways that the rotating away of proton decay could happen: 
(a) The ``heavy fermions" might be the heavier flavors of the Standard
Model ($\tau$, $c$, $b$, $t$), or (b) they might be new, non-Standard Model fermions.

If only the Standard Model quarks and leptons exist, it was shown already in \cite{MinSU5} that proton decay cannot be rotated away in $SU(5)$. If the gauge group is flipped $SU(5)$ \cite{FlippedSU5}, however, it was shown in \cite{DorsnerPerez} that proton decay can be rotated away, although it requires an artificial choice of unification mixing angles.
On the other hand, if there exist new fermions, then it is possible to rotate away  
proton decay even in $SU(5)$. An elegant (though no longer viable) model in which this happens was proposed in \cite{SegreWeldon}. The idea was to impose a conserved abelian charge that absolutely forbids proton decay. This required a doubling of the fermion content of the model. The new, heavy fermions were partnered with the known quarks and leptons in $SU(5)$ multiplets in such a way that proton decay could not occur.

The purpose of this note is to point out that dimension-6 proton-decay 
operators can be rotated 
away in a simple manner in unification schemes based on flipped $SU(5)$ \cite{FlippedSU5}, or
larger unitary groups in which flipped $SU(5)$ is embedded \cite{OtherFlipped}, if there exist extra 
${\bf 5} + \overline{{\bf 5}}$ fermions. The limits in which the proton decay is
completely rotated away are not artificial, but correspond to points in parameter space, where the coefficient of a certain type of term in the Lagrangian becomes small.
We shall first show this in flipped 
$SU(5)$ and then in flipped $SU(6) \times SU(2)$.

Consider a model in which the gauge group is $SU(5) \times U(1)_X$, with each
family consisting of 

\begin{equation}
[{\bf 10}^{(1)} + \overline{{\bf 5}}^{(-3)} + {\bf 1}^{(5)}]
+ [{\bf 5}^{(-2)} + \overline{{\bf 5}}^{(2)}]. 
\end{equation}

\noindent
If $SU(5) \times U(1)_X$ were
embedded in $SO(10)$, the multiplets in the first square 
brackets would make up a ${\bf 16}$, while those in the second square brackets would
make up a ${\bf 10}$. (However, if the model is embedded in $SO(10)$, it turns out
that proton decay cannot be ``rotated away" in the manner that we shall discuss.) 
Let us denote by $Y_5/2$ the $SU(5)$ generator $diag(\frac{1}{2}, \frac{1}{2},
-\frac{1}{3}, -\frac{1}{3}, -\frac{1}{3})$ and by $Y/2$
the weak hypercharge generator of the Standard Model. Then, as is well known, the
flipped embedding of the Standard Model group in $SU(5) \times U(1)_X$ gives
$Y/2 = \frac{1}{5}(- Y_5/2 + X)$. Let us, for the sake of simplicity, consider a model with a single
family. The generalization to three families is trivial. The $SU(5) \times U(1)_X$ 
fermion multiplets then decompose as follows:

\begin{equation}
\begin{array}{ccccccccc}
{\bf 10}^{(1)} & = & \psi^{[\alpha \beta]} & = & \psi^{[12]} & = & N^c & = & (1,1, 0) \\
& & & = & \psi^{[1a]} & = & u & = & (3,2,\frac{1}{6}) \\
& & & = & \psi^{[2a]} & = & d & = & (3,2,\frac{1}{6}) \\   
& & & = & \psi^{[ab]} & = & d^{' c} & = & (\overline{3},1, \frac{1}{3}) \\
& & & & & & & & \\
\overline{{\bf 5}}^{(-3)} & = & \psi_{\alpha} & = & \psi_1 & = & e' & = & (1,2,-\frac{1}{2}) \\
& & & = & \psi_2 & = & \nu' & = & (1,2,-\frac{1}{2}) \\
& & & = & \psi_a & = & u^c & = & (\overline{3},1,-\frac{2}{3}) \\   
& & & & & & & & \\
{\bf 1}^{(5)} & &  & = & \psi & = & e^c & = & (1, 1, +1)  \\
& & & & & & & & \\
\overline{{\bf 5}}^{(2)} & = & \tilde{\psi}_{\alpha} & = & \tilde{\psi}_1 
& = & -\overline{N} & = & (1,2, +\frac{1}{2}) \\
& & & = & \tilde{\psi}_2 & = & \overline{E} & = & (1,2,+\frac{1}{2}) \\
& & & = & \tilde{\psi}_a & = & D^{'c} & = & (\overline{3},1, \frac{1}{3}) \\ 
& & & & & & & & \\
{\bf 5}^{(-2)} & = & \tilde{\psi}^{\alpha} & = & \tilde{\psi}^1 & = & -N' & = & (1,2,-\frac{1}{2}) \\
& & & = & \tilde{\psi}^2 & = & E' & = & (1,2,-\frac{1}{2}) \\
& & & = & \tilde{\psi}^a & = & \overline{D^c} & = & (3,1,-\frac{1}{3}). 
\end{array}
\end{equation}

\noindent The primes on certain multiplets refer to the fact that they are not
mass eigenstates. As we shall see, $d^{'c}$ and $D^{'c}$ mix, as do
$e'$ and $E'$, and $\nu'$ and $N'$.

Several Higgs multiplets are required to do the symmetry breaking and
give mass to the fermions. A ${\bf 10}^{(1)}_h$ of Higgs fields, which we denote
$\Omega^{[\alpha \beta]}$, obtains a superheavy VEV: $\langle \Omega^{[12]} \rangle
= \Omega$. This breaks $SU(5) \times U(1)_X$ down to the Standard Model
group in the well-known way. Two Higgs multiplets are needed to do 
the breaking of the electroweak
group and give weak-scale masses to the quarks and leptons, namely
a ${\bf 5}^{(-2)}_h$ and a ${\bf 5}^{(3)}_h$, which we denote respectively 
as $h^{\alpha}$
and $H^{\alpha}$. These have the nonzero VEVs $\langle h^1 \rangle = v$ and
$\langle H^2 \rangle = V$. 

There are three terms that give fermions superheavy masses:

\begin{equation}
\begin{array}{l}
M \overline{{\bf 5}}^{(2)} {\bf 5}^{(-2)}  
\; + \; Y_d {\bf 10}^{(1)} {\bf 5}^{(-2)} \langle {\bf 10}^{(1)}_h \rangle
\; + \; Y_L  \overline{{\bf 5}}^{(-3)} \overline{{\bf 5}}^{(2)} 
\langle {\bf 10}^{(1)}_h \rangle \\ \\ 
= \; M \; \tilde{\psi}_{\alpha} \tilde{\psi}^{\alpha} \; + \; 
Y_d \; \psi^{\alpha \beta} \tilde{\psi}^{\gamma} \langle \Omega^{\delta \eta}
\rangle \epsilon_{\alpha \beta \gamma \delta \eta}
\; + \; Y_L \psi_{\alpha} \tilde{\psi}_{\beta} \langle \Omega^{\alpha \beta}
\rangle \\ \\
\longrightarrow  
\; M \; [\overline{D^c} \; D^{'c} + E' \; \overline{E} + N' \; \overline{N}] \; + \; 
Y_d \; (d^{'c} \; \overline{D^c}) \Omega 
\; + \; Y_L \; [e' \overline{E} + \nu' \overline{N}]  \Omega 
\end{array}
\end{equation}    

\noindent Collecting terms, this gives $\overline{D^c}(M \; D^{'c} + 
Y_d \Omega  \; d^{'c})$,
$\overline{E} \; (M \; E' + Y_L \Omega \; e')$, and 
$\overline{N} \; (M \; N' + Y_L \Omega \; \nu')$. Defining
$\tan \theta_d = Y_d \Omega/M$ and $\tan \theta_L = Y_L \Omega/M$, 
we see that these terms give superheavy masses to the linear
combinations $D^c \equiv \cos \theta_d D^{'c} + \sin \theta_d d^{'c}$,
$E \equiv \cos \theta_L E' + \sin \theta_L e'$, and 
$N \equiv \cos \theta_L N' + \sin \theta_L \nu'$. The orthogonal
linear combinations remain light (and acquire weak-scale masses from
the Yukawa terms to be discussed below). We denote these light
linear combinations by $d^c \equiv - \sin \theta_d D^{'c} + \cos \theta_d d^{'c}$,
$e \equiv -\sin \theta_L E' + \cos \theta_L e'$, and 
$\nu \equiv -\sin \theta_L N' + \cos \theta_L \nu'$. 

The $SU(5) \times U(1)_X$ fermion multiplets written in terms of the
light and heavy mass eigenstates thus become

\begin{equation}
\begin{array}{l}
{\bf 10}^{(1)} = \left( N^c, \; \left( \begin{array}{c} u \\ d \end{array}
\right), \; \; [\cos \theta_d d^c + \sin \theta_d D^c] \right), \\ \\
\overline{{\bf 5}}^{(-3)} =  \left( \left( \begin{array}{c}  \cos \theta_L
 e + \sin \theta_L E \\ \cos \theta_L \nu + \sin \theta_L N  \end{array}
\right), \; u^c \right), \\ \\ {\bf 1}^{(5)} = (e^c), 
\\ \\
\overline{{\bf 5}}^{(2)} = \left( \left( \begin{array}{l} - \overline{N}
\\ \overline{E} \end{array} \right), \; [-\sin \theta_d d^c + \cos \theta_d
D^c] \right), \\ \\
{\bf 5}^{(-2)} = \left( \left( \begin{array}{l} \sin \theta_L \nu - 
\cos \theta_L N 
\\ -\sin \theta_L e + \cos \theta_L E  \end{array} \right), \; 
\overline{D^c} \right).
\end{array}
\end{equation}

\noindent One sees that in the limit $\cos \theta_d = 0$ and 
$\cos \theta_L = 0$, the Standard Model fermion 
multiplets $(u,d)$, $u^c$, $d^c$, $(\nu, e)$, and $e^c$
are all in different $SU(5)$ multiplets. Thus the superheavy $SU(5)$
gauge bosons cannot cause transitions among the light fermions, 
but always turn a light fermion into a heavy one. Therefore,
in this limit, there is no proton decay mediated by superheavy
gauge bosons. This limit corresponds to $M \ll Y_d \Omega, Y_L \Omega$.

Note that if this model were embedded in $SO(10)$, then there would be
gauge bosons that mediated proton decay even in the limit
$\cos \theta_d = \cos \theta_L = 0$. In particular, in $SO(10)$
there are gauge bosons 
that transform as ${\bf 10}^{(-4)}$ under $SU(5) \times U(1)_X$ and 
that make transitions ${\bf 1}^{(5)} \rightarrow {\bf 10}^{(1)}$ and 
${\bf 10}^{(1)} \rightarrow \overline{{\bf 5}}^{(-3)}$. One of
these couples to $(\overline{u} \gamma^{\mu} e^c)$ and  
$(\overline{u^c} \gamma^{\mu} d)$ and thus gives the dimension-6  
operator $(\overline{u} \gamma^{\mu} e^c) 
(\overline{d} \gamma_{\mu} u^c)$, which violates baryon number
and contains only light fermions and thus produces proton decay.

In the limit $\cos \theta_d = \cos \theta_L = 0$, proton decay
mediated by colored scalars can also be completely suppressed,
as we now show.
In this limit, the mass of the $u$ quark must come from
$\lambda_u {\bf 10}^{(1)} \overline{{\bf 5}}^{(-3)}
\langle \overline{{\bf 5}}^{(2)}_h \rangle =
\lambda_u \; \psi^{[1a]} \psi_a \langle h^*_1 \rangle$. The 
$d$ quark mass must come from 
$\lambda_d {\bf 10}^{(1)} \overline{{\bf 5}}^{(2)}
\langle \overline{{\bf 5}}^{(-3)}_h \rangle =
\lambda_u \; \psi^{[2a]} \tilde{\psi}_a \langle H^*_2 \rangle$. 
And the mass of the $e$ must come from 
$\lambda_{\ell} {\bf 5}^{(-2)} {\bf 1}^{(5)}
\langle \overline{{\bf 5}}^{(-3)}_h \rangle =
\lambda_{\ell} \; \tilde{\psi}^2 \psi \langle H^*_2 \rangle$.
It is easily seen that the interactions of the colored scalars
$h^a$ and $H^a$ produced by these terms only couple a light
fermion to a heavy fermion in the limit
$\cos \theta_d = \cos \theta_L = 0$. Therefore, they do not
produce proton decay. Of course, there could be 
other Yukawa terms that do produce proton decay, but they are not 
needed to give mass to the light fermions.

As we have seen, if flipped $SU(5)$ is embedded in $SO(10)$, it is no 
longer possible to rotate away all the dimension-6 proton-decay operators.
However, they can be rotated away if flipped $SU(5)$ is embedded in
larger unitary groups. We will illustrate this for the case of
$SU(6) \times SU(2)$. (This is a maximal subgroup of $E_6$, but if
$SU(6) \times SU(2)$ is embedded in $E_6$, the dimension-6 proton-decay 
operators can no longer be rotated away.) 

The smallest chiral and anomaly-free set of fermions in 
$SU(6) \times SU(2)$ consists of $(15,1) + (\overline{6}, 2)$, which 
we shall denote $\psi^{[AB]} + \psi^I_A$, where $A,B, ... = 1,2, ..., 6$
are $SU(6)$ indices and $I, J, ... = 1,2$ are $SU(2)$ indices. This set
contains one family. As before, we shall consider a one-family model. The
generalization to three families is trivial.

Two Higgs multiplets with superheavy VEVs are required to break to
the Standard Model group, a $(6, 2)$ that we shall denote $\omega^{A,I}$
and a $(15,1)$ that we shall denote $\Omega^{[AB]}$. The first of
these obtains a VEV $\langle \omega^{6,1} \rangle = \omega$, which has he effect of
breaking $SU(6) \times SU(2)$ down to $SU(5) \times U(1)_X$, where
$X = \frac{1}{2} T_{35} + 5 I_3$. Here $X$ is the generator of $U(1)_X$,
$T_{35} = diag(1,1,1,1,1,-5)$ is a generator of $SU(6)$, and
$I_3 = diag(\frac{1}{2}, -\frac{1}{2})$ is a generator of $SU(2)$.

It is easy to verify that the fermion multiplets decompose under $SU(5)
\times U(1)_X$ into exactly the set
given in Eq. (1):  $\psi^{[\alpha \beta]} = {\bf 10}^{(1)}$,
$\psi^{[\alpha 6]} = {\bf 5}^{(-2)} \equiv \tilde{\psi}^{\alpha}$, 
$\psi^1_{\alpha} = \overline{{\bf 5}}^{(2)} \equiv \tilde{\psi}_{\alpha}$,
$\psi^1_6 = {\bf 1}^{(5)} \equiv \psi$, $\psi^2_{\alpha} = 
\overline{{\bf 5}}^{(-3)} \equiv \psi_{\alpha}$, and $\psi^2_6 =
{\bf 1}^{(0)}$. The superheavy mass terms in Eq. (3) arise from the 
following Yukawa terms:

\begin{equation}
\begin{array}{l}
Y \psi^{[AB]} \psi_A^I \langle \omega_{A,I} \rangle =
Y \psi^{(\alpha 6)} \psi_{\alpha}^1 \langle \omega_{6,1} \rangle 
= (Y \omega) \tilde{\psi}^{\alpha} \tilde{\psi}_{\alpha}, \\ \\
\frac{1}{4}Y_d \psi^{[AB]} \psi^{[CD]} \langle \Omega^{[EF]} \rangle \epsilon_{ABCDEF}
= Y_d \psi^{[\alpha \beta]} \psi^{[\gamma 6]} \langle \Omega^{[\delta \eta]} \rangle \epsilon_{\alpha \beta \gamma \delta \eta}
= Y_d \psi^{[\alpha \beta]} \tilde{\psi}^{\gamma} \langle \Omega^{[\delta \eta]} \rangle \epsilon_{\alpha \beta \gamma \delta \eta}
\\ \\
\frac{1}{2} Y_L \psi^I_A \psi^J_B \langle \Omega^{[AB]} \rangle \epsilon_{IJ}
= Y_L \psi^1_{\alpha} \psi^2_{\beta} \langle \Omega^{\alpha \beta} \rangle
= Y_L \tilde{\psi}_{\alpha} \psi_{\beta} \langle \Omega^{\alpha \beta} \rangle.
\end{array}
\end{equation}

The analysis that led to Eq. (4) is unchanged, and therefore the conclusion still
holds that the superheavy gauge bosons of $SU(5) \times U(1)_X$ do not
lead to proton decay in the limit that $\cos \theta_d
= \cos \theta_L = 0$. Now, however, there are additional superheavy gauge bosons
to consider, namely those of $(SU(6) \times SU(2))/(SU(5) \times U(1)_X)$.
One of these, an $SU(6)$ gauge boson that can be denoted $W^6_a$, has the following couplings involving only light fermions:
$(\overline{\psi^{i6}} \slashed{W}^6_a 
\psi^{ia}) + (\overline{\psi^1_a} \slashed{W}^6_a \psi^1_6)$, which can be written
$(\overline{L} \slashed{W} 
Q) + (\overline{d^c} \slashed{W} e^c)$. In both of these terms $W^6_a$ acts
as though it has $B = -1/3$ and $L = 1$. Therefore, its exchange between light 
fermions does not violate baryon and lepton number. 

There is also a gauge boson in $SU(2)$ that violates baryon number, but it is
easy to see that its couplings always involve the heavy fermions, so it too
does not lead to proton decay. Finally, one can show that the Yukawa terms that
are needed to give weak-scale mass to the light fermions do not couple colored
Higgs fields in such a way as to produce proton decay, as they always couple
light fermions to heavy fermions, as we saw in the flipped $SU(5)$ example.

\end{document}